# Calcium-Decorated Carbon Nanotubes for High-Capacity Hydrogen Storage

Hoonkyung Lee,[1,2] Jisoon Ihm,[3] Marvin L. Cohen,[1,2] and Steven G. Louie[1,2,*]

[1]*Department of Physics, University of California, Berkeley, California 94720, USA*

[2]*Materials Science Division, Lawrence Berkeley National Laboratory, Berkeley, California 94720, USA*

[3]*Department of Physics and Astronomy, Seoul National University, Seoul, 151-747, Korea*

[*]Corresponding author: sglouie@berkeley.edu

## ABSTRACT

Using first-principles calculations, we perform a search for high-capacity hydrogen storage media based on individually dispersed calcium atoms on doped or defective carbon nanotubes. We find that up to six $H_2$ molecules can bind to a Ca atom each with a desirable binding energy of ~0.2 eV/$H_2$. The hybridization of the empty Ca 3$d$ orbitals with the $H_2$ σ orbitals contributes to the $H_2$ binding, and Ca clustering is suppressed by preferential binding of Ca atoms to doped boron and defect sites dispersed on carbon nanotubes. We also show that individual Ca-decorated B-doped CNTs with a concentration of ~6 at. % B doping can reach the gravimetric capacity of ~5 wt % hydrogen.

Hydrogen storage at appropriate density in solid-state materials can be an essential requirement for the development of hydrogen fuel-cell powered vehicles. However, feasible candidate materials are scarce [1,2]. Suggestions for hydrogen storage in carbon nanotubes (CNTs) and other nanostructured materials have been made because of the possibility of reversibility, fast kinetics, and high capacity (large surface area) [3]. Many of the experimental studies were done on pristine organic materials and metal-organic frameworks [3-6]; however, it has been found that the hydrogen storage capacity in these systems greatly diminishes at room temperature and ambient pressure [6,7]. This is attributed to the weak interaction between hydrogen molecules and the materials by physisorption (~0.04 eV) [7]. More recent theoretical studies have been devoted to finding and designing materials which enhance the interactions to the desirable binding energy of $0.2-0.6$ eV [8-17].

In recent years, studies of early transition metal (i.e., Sc, Ti, and V) decorated carbon nanotubes, fullerenes, and polymers have attracted much attention as possible systems for hydrogen storage applications at room temperature and ambient pressure [13-15]. It was expected that these materials not only enhance the interaction between the transition metal element and $H_2$, up to ~$0.2-0.6$ eV via the Kubas interaction (hybridization of the $d$ states with the $H_2$ states) [18], but also meet the U.S. Department of Energy (DOE) goal of obtaining a gravimetric capacity of 9 wt % by the year 2015 [19]. Recently, Antonelli's [20] and Shivaram's [21] experimental groups have shown progress in enhancing the interaction of $H_2$ in reducible mesoporous Ti oxides and Ti-ethylene complexes, which are, unfortunately, still short of the needed interaction in the theoretically suggested structures for room temperature hydrogen storage. Another issue is that, based on energy consideration, transition metal atoms generally prefer being clustered to being individually dispersed on nanomaterials because of their large cohesive energy (i.e., ~4 eV). Clustering changes significantly the adsorption nature of the $H_2$ molecules [22, 23], and this is a major challenge when trying to produce metal-decorated hydrogen storage nanomaterials.

Recently, Yoon *et al*. [24] have suggested another material system, calcium-monolayer-coated fullerenes, for high-capacity hydrogen storage. The motivating idea is that $H_2$ molecules adsorb more strongly on Ca-coated fullerenes than on bare fullerenes because of the polarization of the $H_2$ molecules induced by the electric field produced by the Ca atoms. The charge transfer of the two 4$s$ electrons to the fullerenes causes the fullerenes to appear charged [25]. Calcium-coated $C_{60}$ ($Ca_{32}C_{60}$) is calculated to uptake as many as 92 $H_2$ molecules (2.7 $H_2$ per Ca) with a maximum storage capacity that can reach 8.4 wt %.

Here, instead of considering monolayer Ca-coated fullerenes, we study the case of individual Ca atoms dispersed on carbon nanotubes. Different physical mechanisms from those in Ref. 24 are anticipated. We use Ca as in the transition metal atoms as attractors of $H_2$ molecules, so that the number of adsorbed $H_2$ molecules per Ca is increased via the Kubas interaction. Furthermore, since the cohesive energy of calcium (1.84 eV) is much smaller than that of the transition metals (~4 eV), we expect that the clustering problems in metal-decorated nanostructured materials might be overcome by calcium.

To achieve our goal, we conduct a systematic search for high-capacity hydrogen storage media consisting of individual calcium atoms decorated on doped or defective carbon nanotubes. We find that up to six $H_2$ molecules are attached per Ca with a binding energy of ~0.2 eV/$H_2$ and that the hybridization of the empty Ca 3$d$ orbitals with the $H_2$ σ orbitals contributes to the $H_2$ adsorption. Our calculations show that the Ca atoms may be individually dispersed on B-doped and defective carbon nanotubes without Ca clustering. We also show that the hydrogen storage capacity can reach 5 wt % on 6 at. % B-doped CNTs.

All our calculations were carried out using a first-principles method based on density functional theory as implemented in the Vienna ab-initio simulation package (VASP) with a projector-augmented-wave (PAW) basis set [26,27]. The exchange-correlation energy functional of Perdew and Wang in the generalized gradient approximation (GGA) was used [28], and the kinetic energy cutoff was taken to be 26 ryd. The optimized atomic positions were obtained by relaxation until the Hellmann-Feynman force on each atom was less than 0.01 eV/Å. Supercell calculations were employed throughout where the carbon atoms on adjacent nanotubes are separated by over 10 Å.

To search for high-capacity hydrogen-storage nanostructures consisting of combination of carbon nanotubes and calcium, we first consider a pristine (7,7) CNT and the case of nitrogen substitutional doping, boron substitutional doping, Stone Wales (SW) defect, mono-vacancy (MV), and di-vancancy (DV) on a (7,7) CNT to examine the local structure of the Ca attachment. A single calcium atom is attached to each of the above six structures, and the calculated binding energies of Ca are presented in Table I. We then calculate the binding energy of $H_2$ molecules on Ca in these six structures as a function of the number of $H_2$ molecules adsorbed. Figure 1 shows the optimized geometries for the configuration with maximally adsorbed number of $H_2$ molecules on a single Ca atom for the 6 structures considered. The calculated binding energies of the $H_2$ molecules using the GGA are also presented in Table I. As in previous studies of binding energies of weak bonds, the calculated values with the local density approximation (LDA) are approximately twice as much as those with the GGA, and the correct values are expected to be somewhere in between the two approximations. Up to five (or six) $H_2$ molecules are adsorbed on a Ca atom, and the distance between the Ca atom and the $H_2$ molecule is ~2.45 Å. The bond length of $H_2$ is slightly elongated from 0.75 Å of the isolated molecule to ~0.77 Å.

We investigate whether in the present case the unoccupied Ca $3d$ levels might contribute to the adsorption of the $H_2$ molecules via the hybridization of the $d$ states with the $H_2$ states as in the case of $H_2$ binding to transition metal atoms [18]. Figures 2(a) and 2(b) display the projected density of states (PDOS) for the $H_2$ molecules and the Ca $d$ orbitals when four $H_2$ molecules are adsorbed on a Ca attached on a pristine and B-doped CNTs, respectively. Hybridization of the Ca $3d$ orbitals with the $H_2$ $\sigma$ state at $\sim-9$ eV, and hybridization of the Ca $3d$ orbitals with the $H_2$ $\sigma^*$ orbitals at $\sim-0.1$ eV are seen in Fig. 2(a). Similar results are obtained for the hybridization of the Ca $3d$ states with the $H_2$ $\sigma$ and $\sigma^*$ orbitals in B-doped CNT as shown in Fig. 2(b). These hybridizations are analogous to those found in $H_2$ adsorption on transition metal atoms in the literature [13,14]. For comparison, we show in Fig. 2(c) the PDOS for the H$_2$ molecules and scandium (Sc) when four H$_2$ molecules are adsorbed on a Sc atom attached on to a CNT (with a binding energy of 0.51 eV/H$_2$). It has been shown theoretically that the binding mechanism of H$_2$ to transition metals arises basically from the hybridization of the *d* levels with the H$_2$ states [13]. In Fig. 2(c), the peaks center around $-10 \sim -8$ eV correspond to the hybridization of the Sc 3*d* with the H$_2$ σ orbitals, and the peaks center around $-2 \sim 0$ eV correspond to the hybridization of the Sc 3*d* with the H$_2$ σ$^*$ orbitals, which is the same as the Ca case above except for a smaller strength. These hybridizations were not considered in the interpretation of the results in the Yoon *et al.* [24] paper in which the binding mechanism of H$_2$ molecules is attributed to the induced polarization of the H$_2$ molecules by the Ca attached on fullerenes. According to a recent paper [29], H$_2$ molecules adsorb on Li, Na, and K-decorated B$_{80}$ with a binding energy of ~0.1 eV/H$_2$ where the electric field at the position of the H$_2$ molecule is ~$1.5 \times 10^{10}$ V/m, which corresponds to the value (~$2 \times 10^{10}$ V/m) of the field generated by Ca-attached fullerenes. However, the binding energy of the H$_2$ molecules is only about half as large as that on Ca-attached fullerenes. Therefore, our analysis shows that the hybridization of the Ca 3*d* states with the H$_2$ σ orbitals and the polarization of the H$_2$ molecules both contribute to the H$_2$ adsorption on Ca.

We now address the important issue of metal adsorbate clustering. Our calculations show that Ca clustering takes place on pristine and N-doped CNTs because of the small binding energy of Ca to the CNTs (0.88 eV and 0.61 eV, respectively). Figures 3(a) and 3(b) show the optimized geometries for two aggregated Ca and for individually isolated Ca on pristine CNT where the aggregated case is energetically lower by 0.98 eV (per 2 Ca atoms) as compared with the isolated case. Figure 3(c) shows seven H$_2$ molecules put on the aggregated Ca with a distance of ~2 Å between the H$_2$ molecules and the Ca before energy minimization is performed. The aggregation of Ca results in the dissociation of one H$_2$ molecule as displayed in Fig. 3(d), and therefore the number of adsorbed H$_2$ molecules as well as the binding energy of the H$_2$ molecules are reduced. This behavior is similar to clustering effects of transition metals on other nanomaterials [22,23]. In contrast, when two Ca atoms are individually dispersed on a B-doped CNT, the energy is 1.19 eV (per 2 Ca atoms) lower than the aggregated case as shown in Figs. 3(e) and 3(f). We also confirmed that, isolated Ca atoms attached to defects on CNTs with SW, MV, and DV defects are energetically preferred over the aggregated case by 0.92, 1.70, and 0.82 eV (per 2 Ca atoms), respectively. This is ascribed to a significantly larger binding energy of Ca to the defect site than the calculated-Ca dimer energy (~0.3 eV/Ca in vacuum). We find that the binding energy of Ca near the B site on a B-doped CNT monotonously increases as the B concentration increases. At 3 at. % B doping concentration, the binding energy of Ca reaches 3.2 eV which is much larger than the calculated cohesive energy of bulk Ca, 1.87 eV (1.84 eV experimental value). The enhancement of binding was observed in calculations for transition metals on B-doped fullerenes and graphene as well [13,30]. Therefore Ca can be individually dispersed on CNTs with B doping, SW defect, and vacancies. On the other hand, the aggregation of titanium atoms is still favorable for B-doped CNT as shown in Figs. 3(g) and 3(h) even if the binding energy of the Ti atom is enhanced by 3.4 eV per Ti atom as compared to that (2.2 eV per Ti atom) on pristine CNT. This non-aggregation feature of Ca should make it more effective in binding $H_2$ than the transition metal elements proposed previously in the literature [12-14]. We have also calculated the binding energy of Ca in several well-known organic molecules to explore the tendency toward Ca dispersion. The values for the binding energy are 1.30, 0.16, and 0.63 eV per atom for Ca on fullerene ($C_{60}$), benzene, and *trans*-polyacetylene, respectively. Thus, for pristine organic materials, individual dispersion of Ca atoms is difficult in view of the large bulk cohesive energy of Ca.

We have investigated the effects of nanotube chiral index (metallic vs semiconducting) of B-doped CNTs on Ca and $H_2$ adsorption. On a B-doped (11,0) CNT [which is close to the diameter of a (7,7) CNT], the calculated binding energy of Ca is 1.88 eV and the binding energy of $H_2$ molecules are 0.18, 0.17, 0.19, 0.17, 0.16, and 0.15 eV/$H_2$ for 1, 2, 3, 4, 5, and 6 $H_2$ molecules on Ca, respectively, which are almost the same as those on B-doped (7,7) CNT. These results indicate that the binding energy of Ca or $H_2$ has little to do with the chirality of the CNTs. We checked the curvature effects of CNTs on the binding energy of Ca as well. The binding energy of the Ca atom is calculated to be 1.84, 1.74, and 1.67 eV on B-doped (6,6), (5,5), and (4,4) CNTs, respectively, which varies slightly as the diameter of the CNTs varies. We also found that the dopant B atoms prefer being dispersed as opposed to being clustered on carbon nanotubes. Figure 4 shows the fully optimized structure for a maximum hydrogen-storage capacity in (5,5) CNT with a concentration of 6.25 at. % B, corresponding to 4.94 wt % uptake of $H_2$ molecules where the molecular formula may be expressed as $(C_{75}B_5 \cdot 5Ca \cdot 30H_2)_n$ (n is an integer).

Our results hence demonstrate the potential of Ca-decorated carbon nanotubes with B doping or other defects as a hydrogen storage medium. It has been experimentally observed that B doping concentration in carbon nanotubes and fullerenes can be up to 5 and 20 at. %, respectively [31] and that topological defects such as the Stone-Wales defect and mono- and di-vacancies in carbon nanotubes can be created by ion irradiation for various applications as well [32]. These experimental observations show that the Ca-decorated B-doped or defective CNTs we propose here can be synthesized using current technology. Another attractive feature is that Ca is the fifth most abundant element in the earth's crust. For practical storage applications, although a considerable amount of Ca is needed, there should be ample supply.

In conclusion, we have demonstrated the possibility that individually dispersed Ca-decorated carbon nanotubes can serve as a high-capacity hydrogen storage medium. The empty $d$ states of Ca provide an added mechanism for $H_2$ attachment as in transition metal elements. However, unlike the transition metal elements, Ca has a much lower tendency for clustering on doped or defective CNTs. We feel that these systems can be made, and we encourage an experimental search to synthesize these hydrogen storage nanomaterials that may operate at room temperature and ambient pressure.

This research was supported by the National Science Foundation Grant No. DMR07-05941 and by the Director, Office of Science, Office of Basic Energy Sciences, Materials Sciences and Engineering Division, U. S. Department of Energy under Contract No. DE-AC02-05CH11231. Computational resources were provided by NPACI and NERSC. J. Ihm was supported by the Center for Nanotubes and Nanostructured Composites funded by the Korean Government MEST/KOSEF, and the Korean Government MOEHRD, Basic Research Fund No. KRF-2006-341-C000015.## References

[1] L. Schlapbach and A. Züttel, Nature (London) **414**, 353 (2001).

[2] G. W. Crabtree, M. S. Dresselhaus, and M. V. Buchanan, Phys. Today, **57**, No. **12**, 39 (2004).

[3] A. C. Dillon *et al.*, Nature (London) **386**, 377 (1997).

[4] N. L. Rosi *et al.*, Science **300**, 1127 (2003).

[5] S. S. Kaye and J. R. Long, J. Am. Chem. Soc. **127**, 6506 (2005).

[6] C. Liu *et al.*, Science **285**, 127 (1999).

[7] Y. Ye *et al.*, Appl. Phys. Lett. **414**, 343 (2001).

[8] G. Mpourmpakis *et al.*, Nano Lett. **6**, 1581 (2006).

[9] Q. Sun *et al.*, J. Am. Chem. Soc. **128**, 9741 (2006).

[10] K. R. S. Chandrakumar and S. K. Ghosh, Nano Lett. **8**, 13 (2008).

[11] Y.-H. Kim *et al.*, Phys. Rev. Lett. **96**, 016102 (2006).

[12] N. Park *et al.*, J .Am. Chem. Soc. **129**, 8999 (2007).

[13] Y. Zhao *et al.*, Phys. Rev. Lett. **94**, 155504 (2005).

[14] T. Yildirim and S. Ciraci, Phys. Rev. Lett. **94**, 175501 (2005).

[15] H. Lee, W. I. Choi, and J. Ihm, Phys. Rev. Lett. **97**, 056104 (2006).

[16] E. Durgun *et al.*, Phys. Rev. Lett. **97**, 226102 (2006).

[17] S. Meng, E. Kaxiras, and Z. Zhang, Nano Lett. **7**, 663 (2007).

[18] G. J. Kubas, J. Organomet. Chem. **635**, 37 (2001).
8

**Table I**. Calculated binding energy of Ca (eV/Ca) and $H_2$ molecules (eV/$H_2$) for an isolated Ca atom decorated on a pristine (7,7) CNT and on a nitrogen-dopant, a boron-dopant, a Stone Wales-defect, a mono-vacancy, and a di-vacancy on a (7,7) CNT, respectively, as a function of the number of adsorbed $H_2$ molecules.

|        | Ca   | 1 $H_2$ | 2 $H_2$ | 3 $H_2$ | 4 $H_2$ | 5 $H_2$ | 6 $H_2$ |
|--------|------|---------|---------|---------|---------|---------|---------|
| CNT    | 0.88 | 0.21    | 0.21    | 0.21    | 0.21    | 0.19    |         |
| N-CNT  | 0.61 | 0.23    | 0.22    | 0.21    | 0.21    | 0.19    |         |
| B-CNT  | 1.82 | 0.20    | 0.21    | 0.20    | 0.19    | 0.18    | 0.17    |
| SW-CNT | 2.15 | 0.16    | 0.17    | 0.16    | 0.14    | 0.13    |         |
| MV-CNT | 3.30 | 0.16    | 0.16    | 0.14    | 0.13    | 0.12    |         |
| DV-CNT | 2.20 | 0.20    | 0.19    | 0.17    | 0.16    | 0.15    |         |

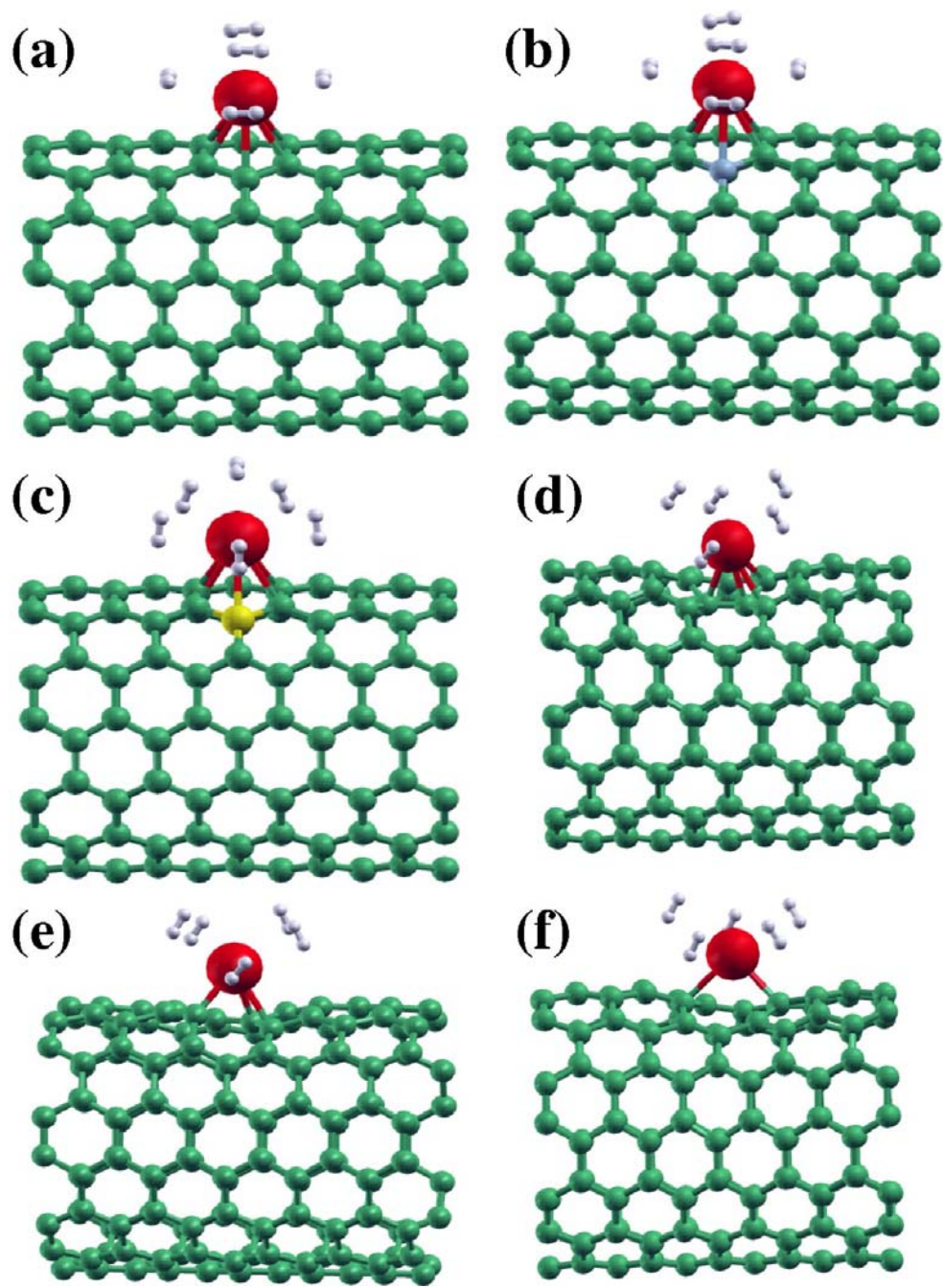

**Figure 1 (Color online):** Optimized atomic geometries of Ca-decorated CNTs with maximum number of H$_2$ molecules attached to the Ca atom. Blue, red, grey, yellow, and white dots indicate the carbon atom, calcium atom, nitrogen atom, boron atom, and hydrogen atom, respectively. (a)−(f) show the geometries for maximally adsorbed H$_2$ molecules to a Ca atom attached to a pristine (7,7) CNT and to a (7,7) CNT with a single N dopant, a single B dopant, a Stone-Wales defect, a mono-vacancy, and a di-vacancy, respectively.

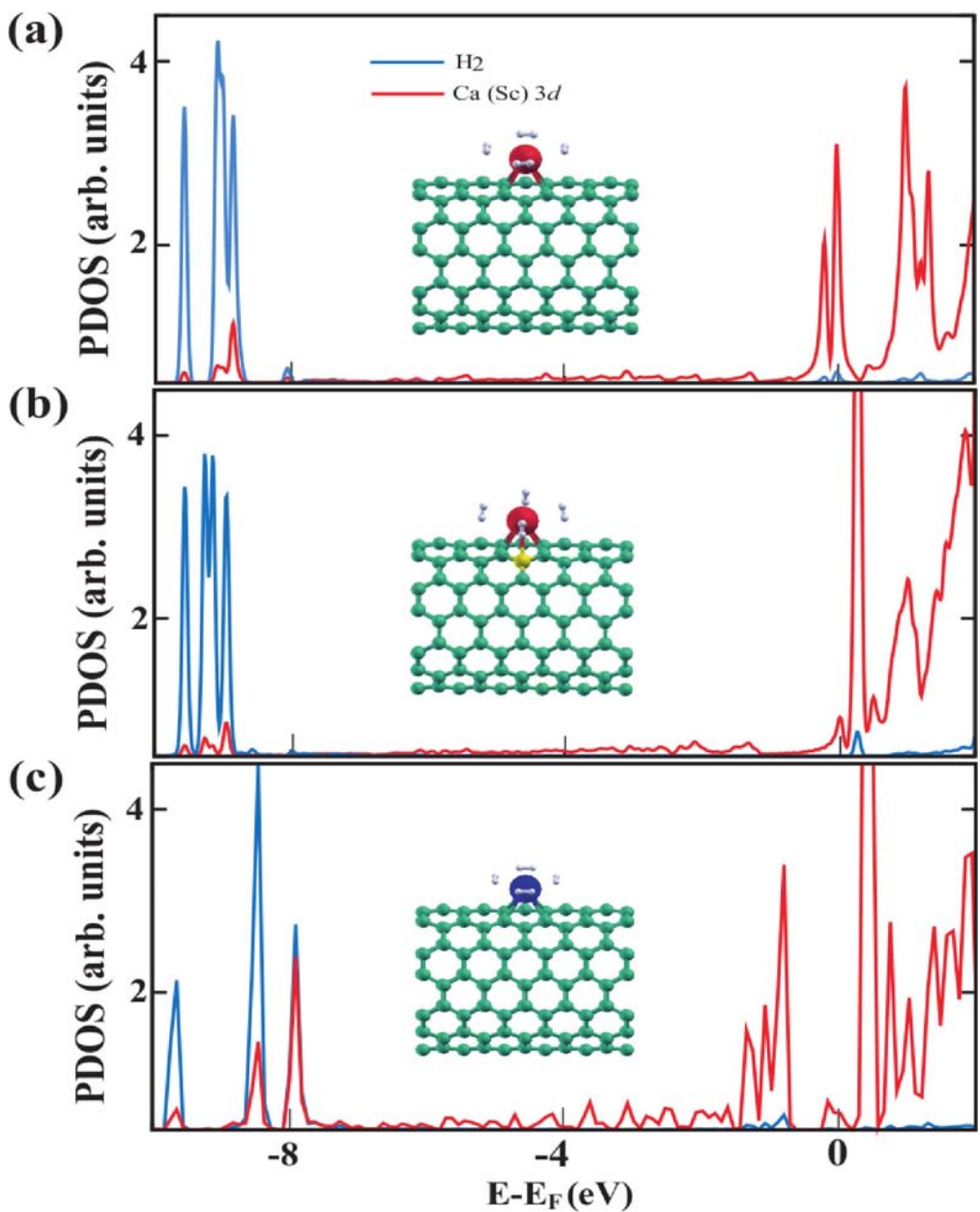

**Figure 2 (Color online):** PDOS of the Ca (Sc) $3d$ orbitals and $H_2$ σ orbitals involved in the adsorption of four $H_2$ molecules on (a) Ca-decorated pristine, (b) Ca-decorated B-doped, and (c) Sc-decorated pristine (7,7) CNTs. The Fermi level is set to zero. The inset of each panel shows the atomic structure with the attachment of four $H_2$ molecules. The blue dot indicates the Sc atom.

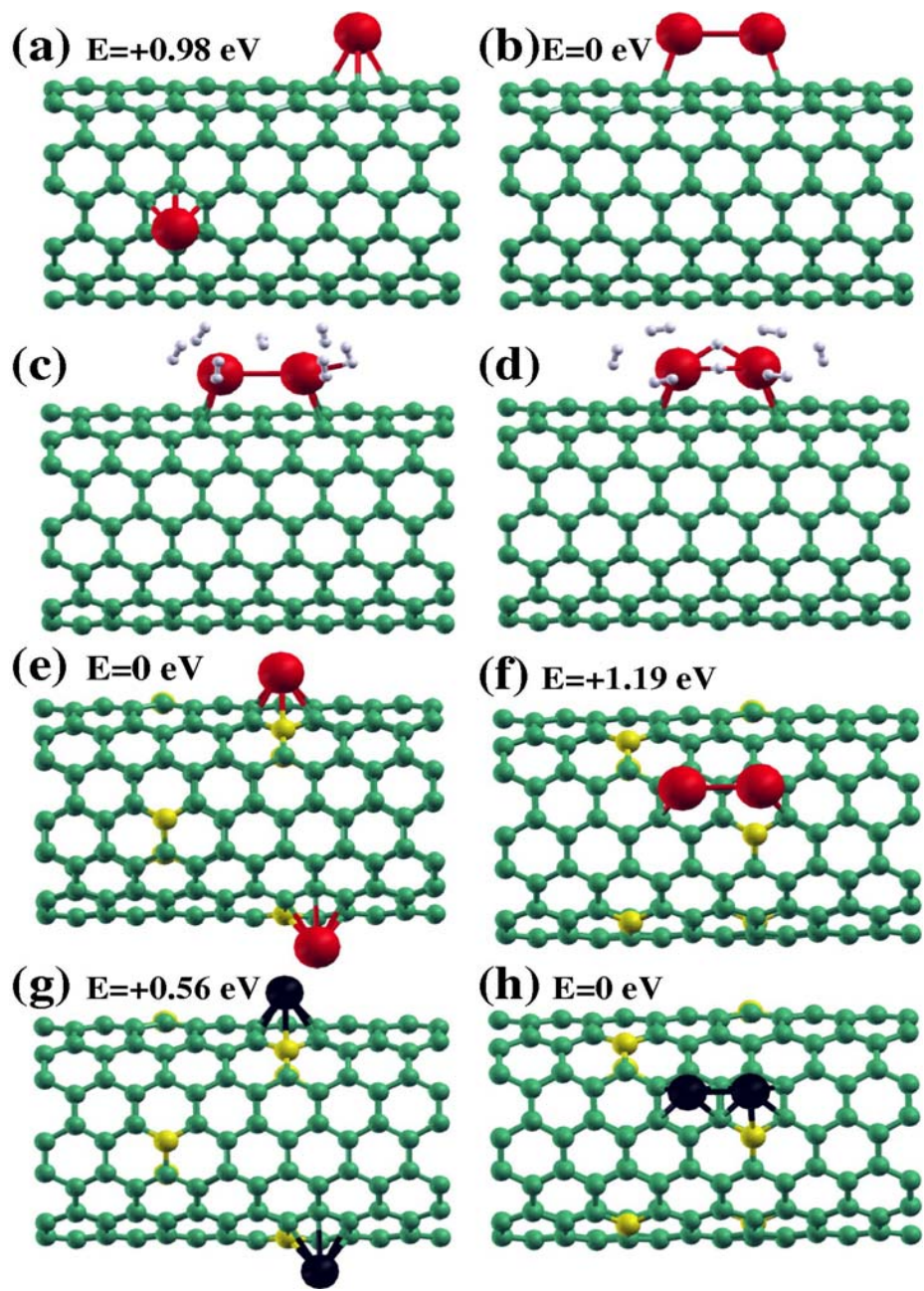

**Figure 3 (Color online):** (a) Two Ca atoms individually attached on a (7,7) pristine CNT. (b) Two Ca atoms aggregated on a pristine (7,7) CNT. (c) Initial geometry of seven $H_2$ molecules on the two aggregated Ca atoms. (d) Geometry of the seven $H_2$ molecules on the two aggregated Ca obtained from energy minimization calculation. (e) Two Ca atoms individually attached on a B-doped (7,7) CNT. (f) Two Ca atoms aggregated on a B-doped (7,7) CNT. (g) Two Ti atoms individually attached on a B-doped (7,7) CNT. Black dots indicate the Ti atoms. (h) Two Ti atoms aggregated on a B-doped (7,7) CNT. The total energy of the lower energy structure between aggregated and unaggregated structure is set to zero.

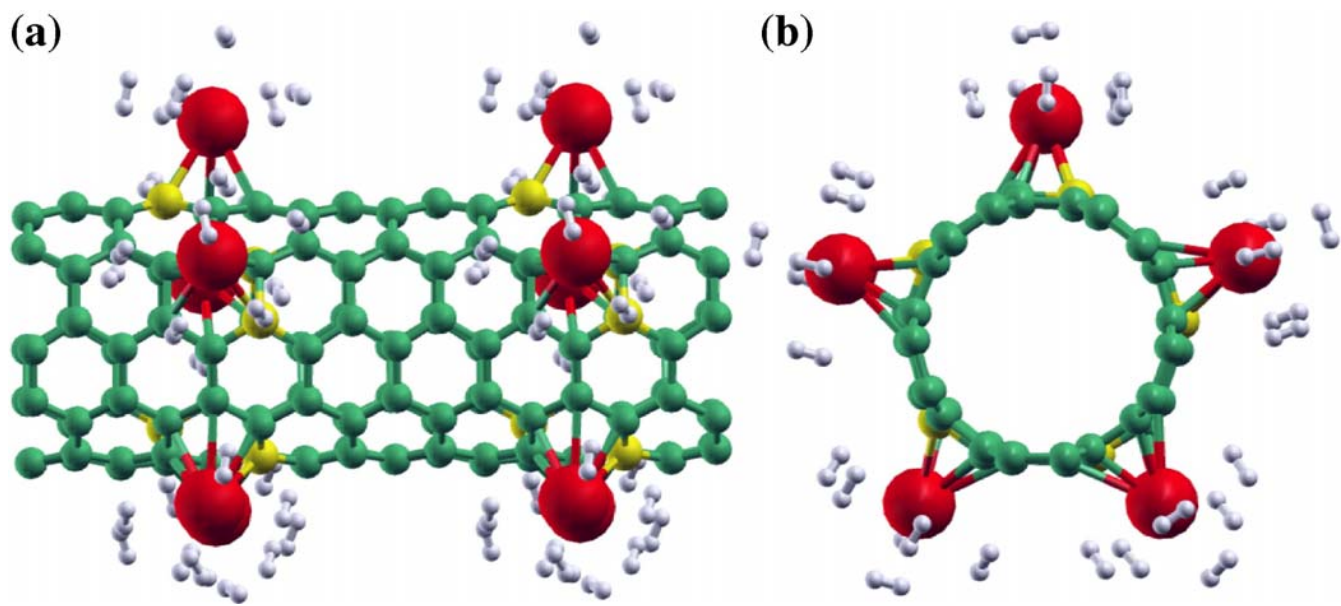

**Figure 4 (Color online):** (a) and (b) show the side view and cross-sectional view for the optimized atomic structure of maximal number of adsorbed $H_2$ molecules for a B-doped (5,5) CNT (6.25 at. % of B), respectively.